\documentclass[
 reprint,
superscriptaddress,
aps,
pra,
 amsmath,amssymb,
floatfix,
]{revtex4-1}
\usepackage{dcolumn}
\usepackage{bm}
\usepackage{geometry}
\usepackage{graphicx} 

\usepackage{float}
\usepackage{multirow}
\usepackage{booktabs} 
\usepackage{array} 
\usepackage{paralist} 
\usepackage{verbatim} 
\usepackage{subfig} 
\usepackage{rotating}
\usepackage{rotating}
\usepackage{tikz}	
\usetikzlibrary{shapes,snakes}
\usetikzlibrary{backgrounds,fit,decorations.pathreplacing}  

\usepackage{fancyhdr} 
\usepackage{threeparttablex}
\usepackage{amsmath, amsthm}
\usepackage{afterpage}
\usepackage[nottoc,notlof,notlot]{tocbibind} 
\usepackage[titles,subfigure]{tocloft} 

\graphicspath{{./figures/}}
\DeclareGraphicsExtensions{.eps}

\usepackage{caption}
\begin{document}

\title{Sensitivity and Entanglement in the Avian Chemical Compass}

\author{Yiteng Zhang}\affiliation{Department of Physics, Purdue University, West Lafayette, IN, 47907 USA}
\author{Gennady P. Berman}\affiliation{Theoretical Division, T-4, MS B-213, Los Alamos National Laboratory, Los Alamos, NM 87545 USA}
\author{Sabre Kais}\email{kais@purdue.edu}
\affiliation{Department of Chemistry, Department of Physics and Birck Nanotechnology Center, Purdue University,
West Lafayette, IN 47907 USA}
\affiliation{Qatar Environment and Energy Research Institute, Qatar Foundation, Doha, Qatar}
\begin{abstract}
The Radical Pair Mechanism can help to explain avian orientation and navigation. Some evidence indicates that the intensity of external magnetic fields plays an important role in avian navigation. In this paper, based on a two-stage strategy, we demonstrate that birds could reasonably detect the directions of geomagnetic fields and gradients of these fields using a yield-based chemical compass that is sensitive enough for navigation. Also, we find that the lifetime of entanglement in this proposed compass is angle-dependent and long enough to allow adequate electron transfer between molecules.
\end{abstract}
\maketitle

$Introduction.$ ---The navigational ability of birds has been a subject of interest for centuries. Every year migrant birds fly hundreds to thousands of miles between their seasonal habitats. Researchers have been trying to explain this astonishing phenomenon for decades. The radical pair mechanism (RPM) is a promising hypothesis to explain this extraordinary phenomenon \cite{rpm0, rad0, rad1, rad2, rpm00}. This RPM was first proposed by Schulten $et$ $al.$ based on the fact that low magnetic fields can alter the rates and the yields of photochemical reactions \cite{rpm0}. This mechanism has been supported by a series of  behavioral experiments \cite{ww1, ww2, ww3, ww4, ww5}, which indicate that the avian compass depends on both inclination and light intensity. Also, the photoreceptor cryptochrome located in the avian retina is involved in this hypothesis \cite{cry1, cry10, cry11, cry12, cry13, cry2, cry3, cry4, cry5}.

Both theoretical and experimental studies have revealed that the internal anisotropic hyperfine interaction is the key factor for a radical-pair-based compass to detect the presence of an external magnetic field and its direction  \cite{rpm0}, \cite{cc, hy0, hyper, sa1, rpm1, kinetics}. The hyperfine interaction has been modeled and simulated in order to understand how to optimize the sensitivity of the chemical compass \cite{hfc1}. Also, the role of quantum coherence and entanglement in the chemical compass has been discussed \cite{et1,et2, et3,et4}. In addition, some research has been done using this theory to design devices to detect the external weak magnetic fields \cite{sa1, sa2, dev1, dev2}.

There are still many aspects of this theory that remain to be studied to improve this mechanism. For instance, the fact that birds can use not only the inclination but also the intensity of geomagnetic fields for navigation has been demonstrated by early behavioral experiments \cite{inten1, inten2}. Also, by comparing the long-distance migration routes and maps of geomagnetic field total intensity \cite{rout, WMM}, one can observe that the migration routes are mostly along the direction of gradient of the magnetic field intensities.

However, little theoretical or computational research has been done to explore this feature of birds' navigation.

In this paper, we study the magnetic field sensitivity of a yield-based chemical compass in birds. As we know, the intensity of geomagnetic field varies from place to place, and the birds can use this information to navigate. We study the magnetic sensitivity of the avian chemical compass to determine how the intensity of geomagnetic field could be utilized in a yield-based compass. Also, we explore the angle-dependence of the magnetic field effect on the yields of radical pairs. In addition, we examine entanglement in the reaction process of the RPM to determine the role that entanglement plays in this mechanism.

\begin{figure}[htpb]
\begin{center}
\includegraphics[width=0.5\textwidth,height=0.35\textwidth]{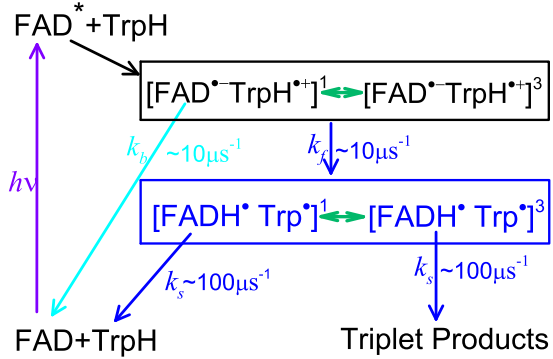}
\end{center}
\caption{\small{The reaction scheme of the radical pair mechanism in cryptochrome. $k_b$ and $k_f$ are the first-order rate constants for recombination of the initial radical pair and formation of the secondary pair from the initial one, respectively. $k_s$ is rate constant for the decay of the secondary pair. The green two-headed arrows indicate the interconversion of the singlet and triplet states of the radical pairs.}}
\label{scheme}
\end{figure}

\begin{figure}[htpb]
\begin{center}
\includegraphics[width=0.5\textwidth,height=0.24\textwidth]{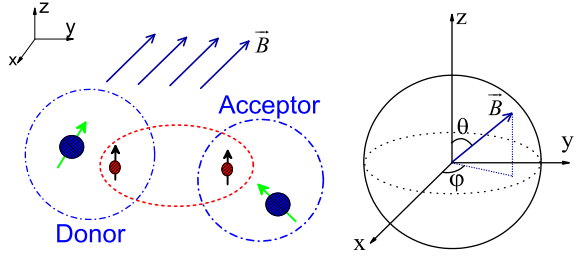}
\end{center}
\caption{\small{Left: A coupled radical pair with neighboring effective nuclear spins (green arrows), in an external magnetic field, $\vec{B}$, (blue arrows). Right: The direction of $\vec{B}$ depicted in the molecular coordinate frame, where the z axis is the z axis of the radical pair \cite{cc}.}}
\label{coor}
\end{figure}

$Model.$ ---Based on the proposed reactions that occur in the photoreceptor molecule, cryptochrome \cite{rpm2}, we consider a two-stage scheme for the radical pair mechanism. The initial radical pair [FAD$\bullet$$^{-}$TrpH$\bullet$$^{+}$] is formed by light-induced electron transfer, followed by the protonation and deprotonation, forming a secondary radical pair [FADH$\bullet$Trp$\bullet$]. This two-stage scheme is shown in Fig. \ref{scheme}. Both radical pairs are affected not only by the external magnetic field but also by their surrounding nuclei. Respectively, the Hamiltonians of the initial and secondary radical pair are,
\begin{equation}
H_1=g \mu_B \sum_{i=1}^2 \vec S_i \cdot \left( \vec B + \widehat A_{1i} \cdot \vec I_i \right), \label{eq1}
\end{equation}
\begin{equation}
H_2=g \mu_B \sum_{i=1}^2 \vec S_i \cdot \left( \vec B + \widehat A_{2i} \cdot \vec I_i \right). \label{eq2}
\end{equation}

In Eq. (\ref{eq1}) and Eq. (\ref{eq2}), $\vec S_i$ is the unpaired electron spin of the radical pairs, and $\vec I_i$ is the nuclear spin of nitrogen in the pairs. For simplicity, we only consider one nitrogen nuclear spin in each molecule. We calculate the hyperfine coupling tensors, $\widehat A_{ij}$, using Gaussian09 with UB3LYP/EPR-II. The hyperfine tensor of the nitrogen nucleus associated with the flavin radical anion FAD$\bullet$$^{-}$ is \textit{diag}$\{$$\widehat A_{11}$$\}$ = $\{$-3.05G, -3.01G, 13.08G$\}$. The hyperfine tensor of the nitrogen nucleus associated with TrpH$\bullet$$^{+}$ is \textit{diag}$\{$$\widehat A_{12}$$\}$ = $\{$5.71G, 5.81G, 27.07G$\}$. The hyperfine tensor of the nitrogen nucleus associated with FADH$\bullet$ is \textit{diag}$\{$$\widehat A_{21}$$\}$ = $\{$-2.80G, -2.69G, 11.53G$\}$. The hyperfine tensor of the nitrogen nucleus associated with Trp$\bullet$ is \textit{diag}$\{$$\widehat A_{22}$$\}$ = $\{$-1.32G, -1.15G, 27.64G$\}$. $\vec B$ is the weak external geomagnetic field. $\vec{B}$ depends on the angles, $\theta$ and $\varphi$, with respect to the reference frame of the immobilized radical pair (Fig. \ref{coor}), i.e., $\vec{B}=B_{0}(\sin\theta\cos\varphi, \sin\theta\sin\varphi, \cos\theta)$. We can choose the x-axis so that the azimuthal angle, $\varphi$, is 0. The constants, $g$ and $\mu_B$, are the $g$-factor and the Bohr magneton of the electron, respectively.

In the two-stage scheme of the RPM, the dynamics of the radical pairs is governed by the following coupled Liouville equations \cite{cc}:

\begin{equation}
\begin{split}
\label{eq3}
\dot{\rho_1}(t)=&-\frac{i}{\hbar}[H_1,\rho_1(t)] \nonumber\\
&-\frac{k_f}{2}\left\{Q^S,\rho_1(t)\right\} -\frac{k_f}{2}\left\{Q^T,\rho_1(t)\right\}\\
&-\frac{k_b}{2}\left\{Q^S,\rho_1(t)\right\},
\end{split}
\end{equation}
\begin{equation}
\begin{split}
\label{eq4}
\dot{\rho_2}(t)=&-\frac{i}{\hbar}[H_2,\rho_2(t)] \nonumber\\
                    &+\frac{k_f}{2}\left\{Q^S,\rho_1(t)\right\}+\frac{k_f}{2}\left\{Q^T,\rho_1(t)\right\}\\
                    &-\frac{k_s}{2}\left\{Q^S,\rho_2(t)\right\}-\frac{k_s}{2}\left\{Q^T,\rho_2(t)\right\},
\end{split}
\end{equation}
where $H_1$ and $H_2$ are the Hamiltonians of the two radical pairs given in Eqs. (\ref{eq1}) and (\ref{eq2}); $\rho_1$ is the density matrix of the initial radical pair, and $\rho_2$ is that of the secondary radical pair; $Q^S$ is the singlet projection operator, $Q^S=|S\rangle\langle S|$, and $Q^T=|T_+\rangle\langle T_+|+|T_0\rangle\langle T_0|+|T_-\rangle\langle T_-|$ is the triplet projection operator, where $|S\rangle$ is the singlet state, and ($|T_+\rangle, |T_0\rangle, |T_-\rangle$) are the triplet states; and all of the decay rates are indicated in Fig. \ref{scheme}. In addition, the initial state of the pair [FAD$\bullet$$^{-}$TrpH$\bullet$$^{+}$] is assumed to be in the singlet state, $\mid$S$\rangle=\frac{1}{\sqrt{2}}(\mid\uparrow\downarrow\rangle-\mid\downarrow\uparrow\rangle)$, while the pair [FADH$\bullet$Trp$\bullet$] is not produced initially. In other words, $\rho_1(0)=\frac{1}{9}\hat{I}_N\otimes {Q}^S$, where the electron spins are in the singlet states, and nuclear spins are in thermal equilibrium, a completely mixed state, which is a 9$\times$9 identity matrix, and $\rho_2(0)=0$.
\begin{figure}[htpb]
\begin{center}
\includegraphics[width=0.5\textwidth,height=0.4\textwidth]{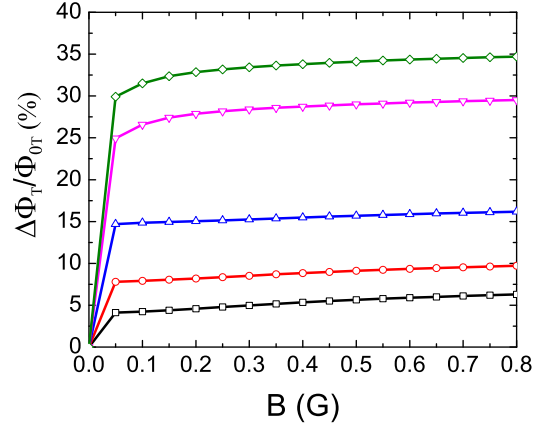}
\end{center}
\caption{\small{Angle dependence of magnetic field effect under different intensities of external fields. The MFE is the ratio of the difference between the triplet yields of the secondary pair under external fields and zero field and the triplet yields under zero field, which is $\Delta\Phi_T/\Phi_{0T}$. This graph depicts the MFE as a function of external fields $\vec{B}$ for five different polar angles, $\theta$, between the electron spin and the magnetic field. $\theta = 0^{\circ}(\text{black}, \Box)$, $30^{\circ}(\text{red}, \circ)$, $60^{\circ}(\text{blue}, \bigtriangleup)$, $85^{\circ}(\text{pink}, \bigtriangledown)$, $90^{\circ}(\text{green}, \diamond)$. The decay rates are shown in Fig. \ref{scheme}.}}
\label{mfe}
\end{figure}

Also, we consider the product formed by the radical pair [FADH$\bullet$Trp$\bullet$] in the triplet state as the signal product, whose yield is defined as $\Phi_T=k_s\int_0^\infty Tr[Q^T\rho(t)] dt$ \cite{yd0}\cite{yd}, where $Q^T=|T\rangle\langle T|$, and $\mid$T$\rangle=|T_+\rangle + | T_0\rangle+ | T_-\rangle$.
\begin{figure}[htpb]
\begin{center}
\includegraphics[width=0.5\textwidth,height=0.4\textwidth]{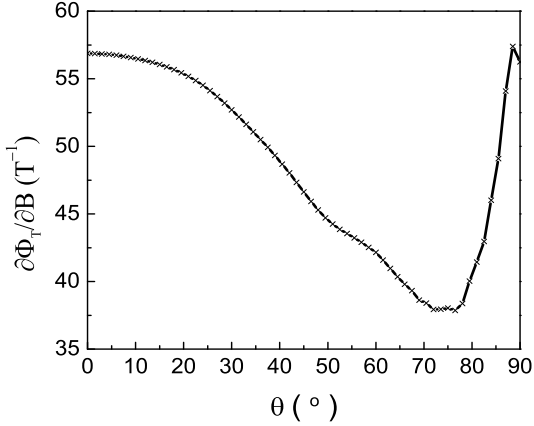}
\end{center}
\caption{\small{The magnetic sensitivity of the chemical compass with respect to angles. The sensitivity is defined as $\partial \Phi_T/\partial B $ in T$^{-1}$, and the angle in degrees. There is a rapid increase in this sensitivity between 80$^{\circ}$ and 90$^{\circ}$.}}
\label{senang}
\end{figure}

$Results~and~Discussion.$ ---First, we look at the magnetic field effect (MFE) on the signal yields. We consider $\Delta\Phi_T/\Phi_{0T}$ as a metric of the magnetic field effect, where $\Phi_{0T}$ is the signal yield at zero field and $\Delta\Phi_T$ is the difference between the signal yields at a given field and $\Phi_{0T}$. From Fig. \ref{mfe}, we can clearly see that the angle plays a very important role in the MFE. As the angle increases, the MFE becomes more and more significant. The yields are barely changed, around 5$\%$, when the angle is 0$^{\circ}$. However, the MFEs are very significant at 85$^{\circ}$ and 90$^{\circ}$, since the yield changes are greater than 25$\%$. The MFE is even greater than 30$\%$ at 90$^{\circ}$. Such a significant effect, possibly exceeding the environmental noise, could induce a physiological reaction in birds, making this yield-based chemical compass feasible for birds when the polar angle is around 90$^{\circ}$. For simplicity of consideration, we assume that the geographical meridians are parallels to the magnetic meridians. Then, the geographical parallels are normal to the meridians. In other words, it is easier for birds to detect the direction of parallels, which indicates the east-west direction, since the geomagnetic field has a significant effect on the yields of radical pair reactions along the parallels. As we know, the climate can mainly be affected by the parallels and the birds immigrate according to the climate. Therefore, this feature may be a result of the natural selection.

Having estimated the role of MFE on the signal yields, we now focus on the magnetic sensitivity of the birds' compass, which is defined as $\partial \Phi_T/\partial B$ (T$^{-1}$) \cite{et4}. From the MFEs for  different polar angles (Fig. \ref{mfe}), we speculate that birds can detect the direction of parallels. The magnetic sensitivity at a magnetic field of 0.5G for various angles (Fig. \ref{senang}) also confirms this conjecture. Fig. \ref{senang} shows that the sensitivities around 0$^{\circ}$ and 90$^{\circ}$ are similar and also larger than for most other angles, which could indicate that the birds can detect the directions of meridians and parallels if they use the intensity of the magnetic field as a clue of navigation, since the yield-based compass is most sensitive along these two directions. Another property that attracted our attention is that the sensitivity's slope is significantly larger between 80$^{\circ}$ and 90$^{\circ}$ than that of the other sections of the curve. Along with the MFEs, this property may imply that it is easier for birds to detect the direction of parallels than that of meridians due to this rapidly increasing sensitivity. Also, we can expect that it is easier for birds to detect the change of the intensity of the external magnetic field when the polar angle is around 90$^{\circ}$ since the yield-based compass is very sensitive to the change of intensities of fields. This capability can enable the birds to immigrate along the direction of the gradient of intensities of the geomagnetic fields.
\begin{figure}[htpb]
\begin{center}
\includegraphics[width=0.5\textwidth,height=0.4\textwidth]{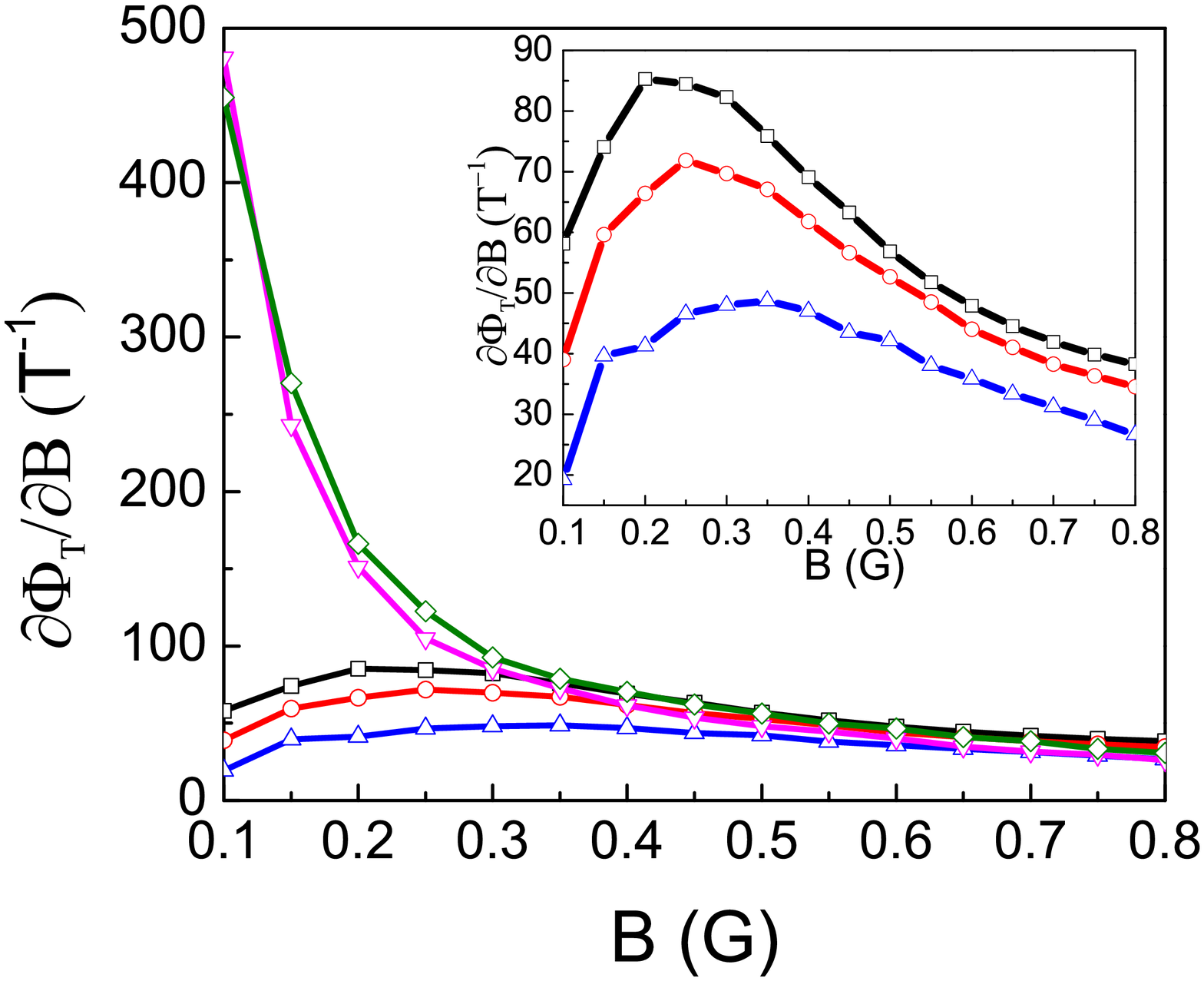}
\end{center}
\caption{\small{Intensity dependence of the magnetic sensitivity. This graph depicts the sensitivity as a function of external magnetic fields $\vec{B}$ under different polar angle $\theta$ between the z axis of the radical pair and magnetic fields, i.e., $\theta = 0^{\circ}(\text{black}, \Box)$, $30^{\circ}(\text{red}, \circ)$, $60^{\circ}(\text{blue}, \bigtriangleup)$, $85^{\circ}(\text{pink}, \bigtriangledown)$, $90^{\circ}(\text{green}, \diamond)$. The inside graph shows the closer view under the angles $0^{\circ}(\text{black}, \Box)$, $30^{\circ}(\text{red}, \circ)$, $60^{\circ}(\text{blue}, \bigtriangleup)$.}}
\label{sen}
\end{figure}

Also, we explore the magnetic sensitivity as a function of the intensities of the magnetic field for several polar angles $\theta$ (Fig. \ref{sen}). Basically, Fig. \ref{sen} shows two patterns of curves. The first pattern is observed for 85$^{\circ}$ and 90$^{\circ}$ for which the sensitivities monotonically decrease as the external fields increase. In this situation, the sensitivities are extremely high in the extraordinary weak magnetic fields, less than 0.25G, which is smaller than the range of the geomagnetic fields, from 0.25G to 0.65G \cite{WMM}. While in the range of geomagnetic fields, the sensitivities fall into the normal range, similar to other angles. The other pattern occurs for 0$^{\circ}$, 30$^{\circ}$ and 60$^{\circ}$, and the sensitivities ascend at first then descend later as the external fields increase. In this situation, the maxima of the curves move rightwards and downwards as the polar angles increase. Combining these two situations (Fig. \ref{sen}), we observe the properties of the chemical compass mentioned before, namely that compass is most magnetically sensitive around 0$^{\circ}$ and 90$^{\circ}$ for geomagnetic fields. However, above 0.35G, all of the sensitivities decrease as the fields' intensities increase. When the intensity of the magnetic field is larger than 0.55G, the magnetic sensitivity becomes very small, less than 40 T$^{-1}$. Therefore, in larger external fields ($\ge$ 0.55G), the chemical compass is less sensitive, and it is difficult for birds to detect the change of the intensities of external fields. As a result, the birds may lose the ability to navigate. Such a feature could explain the experimental phenomenon that the birds are not able to navigate themselves in fields larger than 0.54G \cite{cc}.

\begin{figure}[htpb]
\begin{center}
\includegraphics[width=0.5\textwidth,height=0.4\textwidth]{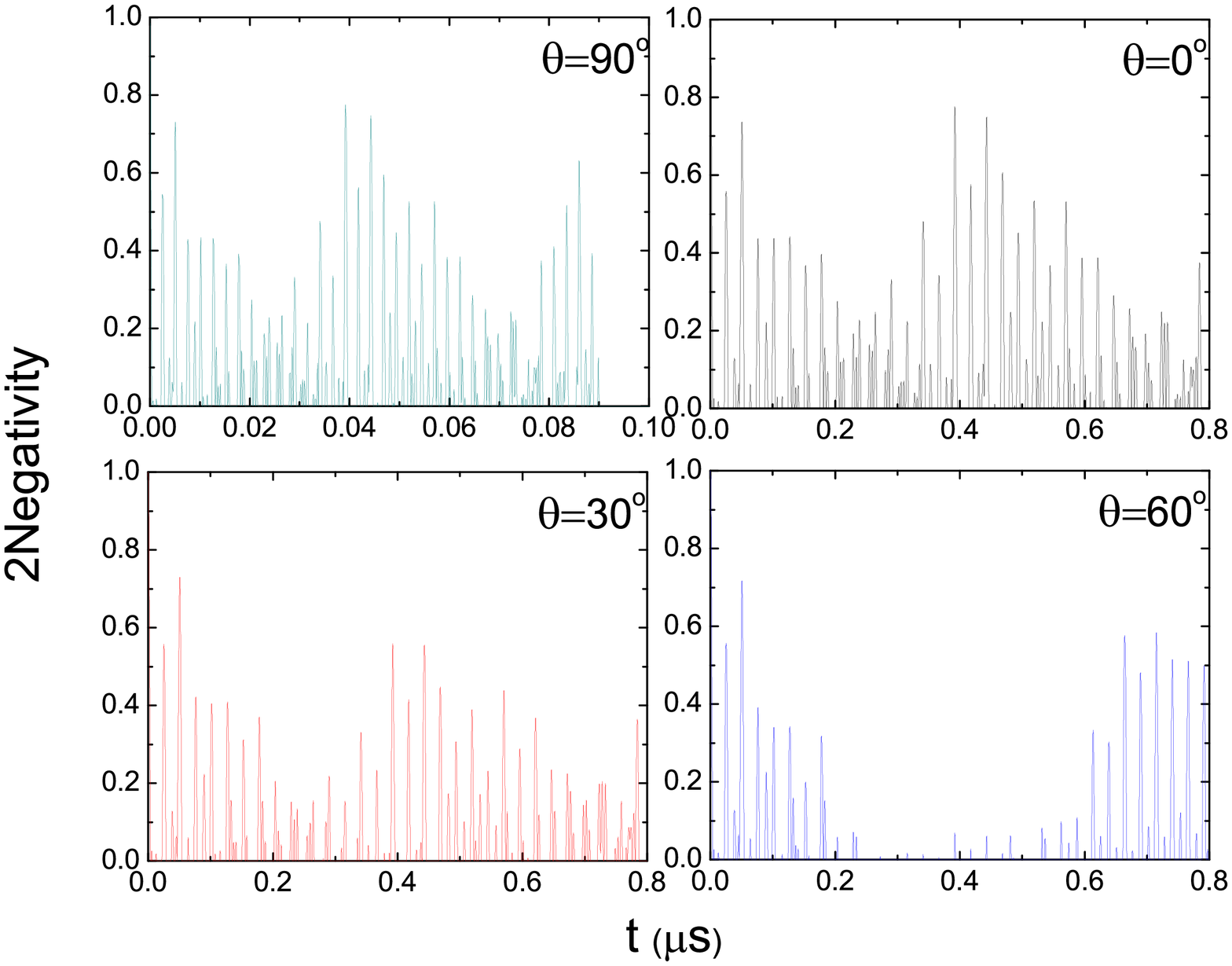}
\end{center}
\caption{\small{Entanglement of the initial radical pair [FAD$\bullet$$^{-}$TrpH$\bullet$$^{+}$] as a function of external fields for four polar angles, $\theta = 0^{\circ}(\text{black}), 30^{\circ}(\text{red}), 60^{\circ}(\text{blue}), 90^{\circ}(\text{green})$. Since the entanglement of he initial pair [FAD$\bullet$$^{-}$TrpH$\bullet$$^{+}$] is compressed within 0.1 $\mu$s, the time scale of the graph for that is from 0 to 0.1 $\mu$s. The other graphs range from 0 to 0.8 $\mu$s. And the entanglement of the initial pair at $0^{\circ}, 30^{\circ}, 60^{\circ}$ differ after 0.3 $\mu$s.}}
\label{entg2}
\end{figure}
Entanglement is believed to play an important role in many systems\cite{kais1,kais2,kais3,kais4} including the chemical compass in birds. However, for the two-stage scheme, the secondary radical pair barely has entanglement between the two unpaired electrons, since the chemical reaction has destroyed the entanglement between them in the preceding radical pair [FAD$\bullet$$^{-}$TrpH$\bullet$$^{+}$]. The unpaired electrons in the initial radical pair show a robust entanglement. Fig. \ref{entg2} shows the entanglement of the initial radical pair [FAD$\bullet$$^{-}$TrpH$\bullet$$^{+}$] for four polar angles, $\theta$. Also, the dynamics of the entanglement is clearly dependent on the angles. However, the entanglements at 0$^{\circ}$, 30$^{\circ}$ and 60$^{\circ}$ are nearly the same for the first 0.1$\mu$s, while the entanglement at 90$^{\circ}$ is very different from others. At 90$^{\circ}$, the entanglement lasts for 0.1$\mu$s, which is long enough for electrons to transfer between different molecules \cite{timescale}.

\begin{figure}[htpb]
\begin{center}
\includegraphics[width=0.5\textwidth,height=0.4\textwidth]{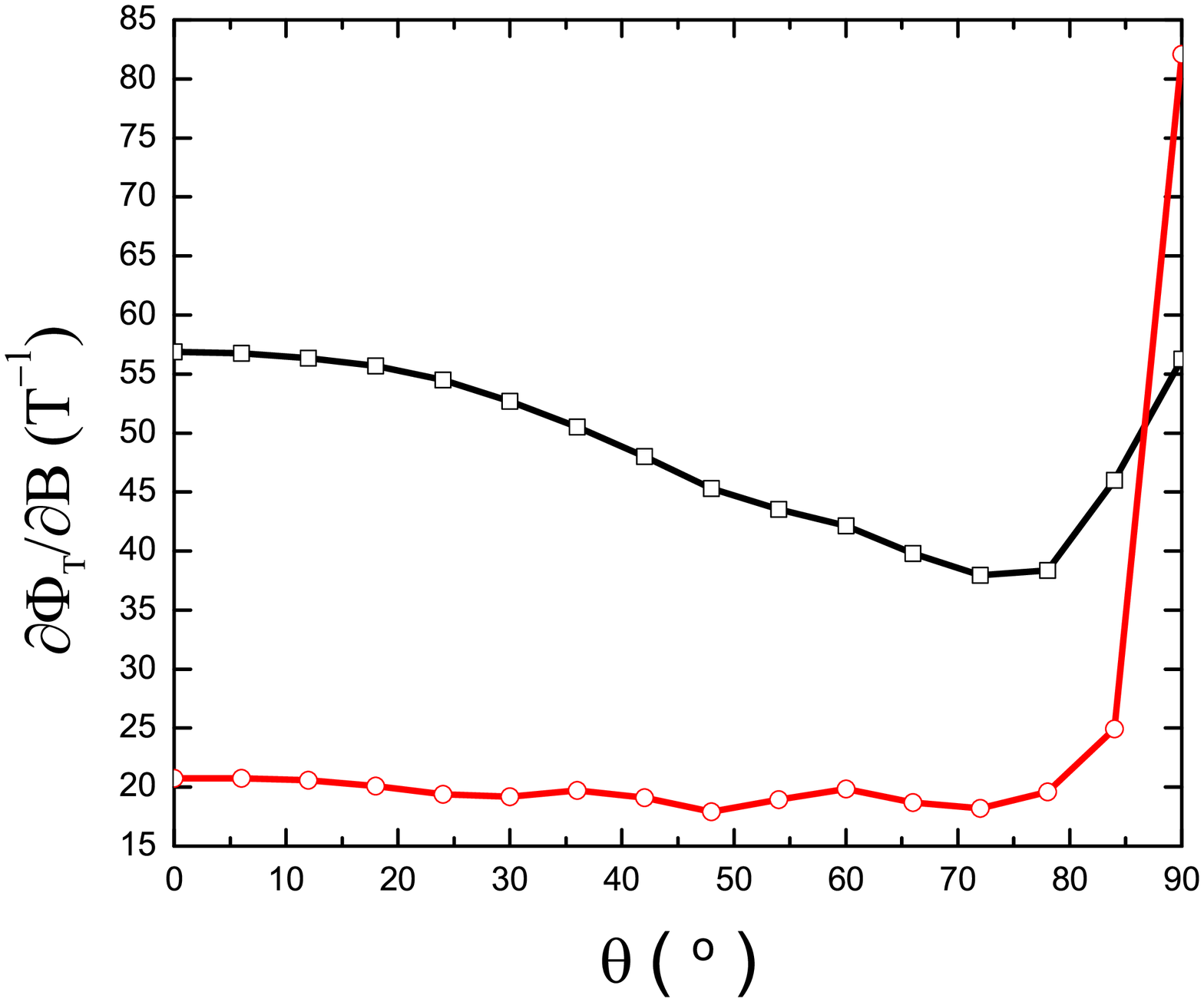}
\end{center}
\caption{\small{The magnetic sensitivity as a function of angles under different hyperfine coupling tensors. The black curve ($\Box$) describes the situation under the hyperfine tensor mentioned in the previous context, while the red one ($\circ$) shows the situation under a different set of hyperfine tensor, \textit{diag}$\{$$\widehat A_{11}$$\}$ = $\{$-0.650G, -0.566G, 17.071G$\}$, \textit{diag}$\{$$\widehat A_{12}$$\}$ = $\{$5.71G, 5.81G, 27.07G$\}$, \textit{diag}$\{$$\widehat A_{21}$$\}$ = $\{$-0.792G, -0.692G, 14.060G$\}$, \textit{diag}$\{$$\widehat A_{22}$$\}$ = $\{$-1.32G, -1.15G, 27.64G$\}$.}}
\label{hfc}
\end{figure}

$Conclusions.$ ---We have studied the role of the intensity of the magnetic field in the birds' navigation. The properties around 90$^{\circ}$, i. e., the direction of parallels, stand out from those at other angles in the aspects of magnetic sensitivity and entanglement. When the birds migrate thousands of miles, they may be able to detect the change of the intensity of geomagnetic fields and the approximate direction of parallels instead of sensing the exact direction. From our simulations, we provide some preliminary justification. We also changed the set of hyperfine tensors to calculation the angle dependence of sensitivity in the field of 0.5G (Fig. \ref{hfc}). The result also shows that the magnetic sensitivity is extraordinarily high around 90$^{\circ}$, although the sensitivity at 0$^{\circ}$ becomes very low, which gives a support to our conjecture that the birds are able to detect the approximate direction of parallels.

This research also demonstrated that birds can use head scans to determine orientation \cite{headscan}. By using head scans, they may adjust the polar angle (Fig. \ref{coor}) to try to find the direction of parallels. After finding the direction of parallels, they can make the chemical compass most sensitive to the change of intensities of geomagnetic fields along such a direction, so that they can fly along the direction of the gradient of the intensities. Therefore, by detecting the directions of parallels and the gradient of the intensities of geomagnetic fields, the birds are able to migrate thousands of miles.

$Acknowledgements.$ ---We would like to thank the NSF Center for Quantum
Information  for Quantum Chemistry (QIQC), Award
No. CHE-1037992, for financial support.

The work by G.P.B. was carried out under the auspices of
the National Nuclear Security Administration of the U.S. Department of Energy
at Los Alamos National Laboratory under Contract No. DE-AC52-06NA25396.

\end{document}